# Knowing oneself through and with AI:

# From self-tracking to chatbots

## Lucy Osler

**Abstract:** This chapter examines how algorithms and artificial intelligence are transforming our practices of self-knowledge, self-understanding, and self-narration. Drawing on frameworks from distributed cognition, I analyse three key domains where AI shapes how and what we come to know about ourselves: self-tracking applications, technologically-distributed autobiographical memories, and narrative co-construction with Large Language Models (LLMs). While self-tracking devices promise enhanced self-knowledge through quantified data, they also impose particular frameworks that can crowd out other forms of self-understanding and promote self-optimization. Digital technologies increasingly serve as repositories for our autobiographical memories and self-narratives, offering benefits such as detailed record-keeping and scaffolding during difficult periods, but also creating vulnerabilities to algorithmic manipulation. Finally, conversational AI introduces new possibilities for interactive narrative construction that mimics interpersonal dialogue. While LLMs can provide valuable support for self-exploration, they also present risks of narrative deference and the construction of self-narratives that are detached from reality.

## Introduction

The relationship between digital technologies and self is a much-discussed topic. A common trope about the internet is that it allows us to be whoever we want to be, providing an open space for self-exploration and self-expression. Indeed, Sherry Turkle suggests that one of the great attractions of going online is that digital spaces offer opportunities "for performing as the self you wanted to be" (Turkle 2017, 158). However, others (including Turkle) have emphasised that engaging with various technologies actively shapes what we can do, what we care about, and transform *who* we are (e.g., Haraway 2013; Ihde 2002; Nguyen 2021a, b; Osler 2024).

In this chapter, I explore various ways in which engaging with AI-driven digital technologies shapes our self-knowledge, self-understanding, and self-narratives.[1] I first examine the increasing reliance on self-tracking devices for obtaining information about oneself in the form of quantified data. I emphasise that these technologies do not furnish us with neutral information, but select and shape the data we are presented with, thus shaping how we interpret and understand ourselves. Second, I look more broadly at the way in which we distribute our self-narratives onto digital technologies. And finally, and perhaps most interestingly, I analyse new possibilities offered by generative AI chatbots for co-constructing our self-narratives. Across each of these sections, I draw attention to various promises and perils that emerge from knowing, understanding, and narrating oneself through and with AI.

**1. Self-tracking apps and the datafied self**

In 2007, Gary Wolf and Kevin Kelly founded the "Quantified Self" network—a network that promotes the monitoring of 'the self' through digital devices such as wearable technology and apps. The tagline of the Quantified Self website is "self-knowledge through numbers".[2] The network discusses various aspects of our lives and selves that can be tracked, including diet, steps, sleep cycles, mood, productivity, calorie intake, phone usage, and more. The network not only considers what self-tracking options may currently be available but also engages with how future technology can be designed to "better help us answer a broader range of questions about ourselves". Examples of wearable devices include: Fitbits, the Apple Watch, and the Oura Ring; examples of self-tracking apps include: MyFitnessPal, Apple Fitness, and Clue. Even apps that are not advertised as self-tracking increasingly come with in-built self-tracking dimensions, for instance Spotify's yearly metrics about your listening habits.

Often self-trackers can provide us with information about ourselves that would not typically be available to us; either because they are hard to keep track of without such devices (e.g., step count) or because they are monitoring bodily functions such as heart rate or eye-movement. On first flush, it seems that such technologies increase the amount of information we can access about ourselves and, thus, are excellent tools for improving our self-knowledge. Indeed, there are many reports of self-tracking applications helping people understand their bodily and affective selves in more depth and detail; testimonies about people using self-tracking technologies to track allergies, understand follow-through for commitments, get insight into exercise regimes and achievements, and more.[2] Self-tracking has received particular popularity in relation to women's health issues, for instance, with many women using apps such as Endo Empowered to track symptoms that have helped them, and their doctors, to diagnose them with conditions such as endometriosis—a condition that is infamously poorly understood and underdiagnosed (Hallström 2024; Richardson-Self & Osler forthcoming). In such cases, individuals not only have access to data about themselves, but this data can be both epistemically and affectively empowering—providing evidence that can be shared, acted upon, and can dispel uncertainty and doubt.[3]

---

[1] Throughout the chapter, I distinguish between self-knowledge (broadly, factual information about oneself), self-understanding (broadly, interpretations of this information, one's actions, and one's sense of who one is), and self-narratives (broadly, the stories we construct about who we are and are becoming). Note that these are deeply interrelated: narratives depend on knowledge and understanding, while simultaneously shaping what knowledge we attend to and how we understand ourselves. Importantly, I remain neutral on the metaphysical question of whether there is a substantial "self" that is the object of such knowledge, understanding, and narration. My analysis concerns the practices through which people form beliefs and narratives about themselves, regardless of whether these practices track a metaphysically robust entity that is "the self".

However, without wanting to dismiss self-tracking *tout court*, it is important to recognise that these devices do not simply furnish us with neutral information about ourselves, they select and shape the data we have access to and, thus, play a role in how we come to understand ourselves. In his paper *The Seduction of Clarity* (2021a), C. Thi. Nguyen considers our tendency to be seduced by a feeling of clarity. Often, we experience an affectively pleasurable "a-ha" moment when we have understood something (Gopnick 1998). Feeling that we have gained understanding can encourage us to cease any further inquiry into the matter at hand and amplifies epistemic confidence. If we can be made to experience this "a-ha" moment, we can be fooled into thinking we have attained genuine comprehension or insight, even when this is, in fact, not the case.

Quantified systems "offer an exaggerated sense of clarity without an accompanying amount of understanding or knowledge" (Nguyen 2021a, 229). The quantified data that self-tracking technologies provide give us a seductively clear picture of our own bodies, habits, and self. With a few swipes of my phone, I am offered information about how far I've walked today, how many calories I've burnt, whether I am a morning person, whether I have a set routine, or if I am meeting my new year's resolution to work harder. However, it often does so at the expense of nuance, context, and other forms of self-understanding.

In using and consulting self-tracking technologies, we accept a quantified framing, or datafication, of our selves. It encourages us to approach self-knowledge and understanding as something that can be conveyed through data and numbers. This has a number of potentially worrying implications. First, self-tracking devices do not simply provide us with data about us, they select what data we are provided with. This is often based on practical considerations such as what the technology can track or what a user can readily input and what can be easily quantified. For instance, fitness apps (roughly) track step counts, as this is something our devices can do with a reasonable degree of accuracy, and, as such, step counts become representative of 'fitness'. In relying on self-tracking apps for self-knowledge, we are, often inadvertently, shifting our understanding of ideas such as wellness, productivity, and resilience to fit with the framing of such devices. Note that this is not necessarily problematic, perhaps there are advantages to such frameworks. Yet, as Nguyen pointedly puts it: "step counts are not the same as health, and citation rates are not the same as wisdom" (2021a, 229).

---

[2] https://quantifiedself.com/show-and-tell/
[3] Health tracking apps can also serve as the basis for developing community-based knowledge production and social connection (see Tempini 2015).

Second, and in a related vein, we might worry that forms of self-knowledge that are difficult to quantify fall out of the picture, becoming less salient in our self-understanding; conceptions such as kindness, patience, and bravery. This disrupts the idea that self-tracking apps straightforwardly provide us with more knowledge about ourselves, suggesting that it might more accurately be described as shifting the way in which we gather self-knowledge about ourselves, potentially crowding out other means.

Third, the "quantified self" is often critiqued for promoting neoliberal values of self-evaluation and self-worth, rendering us more exploitable workers and citizens (Lupton 2016; Moore 2017; Ajana 2017). Take, for instance, how a government might for fiscal reasons encourage its citizens to use self-trackers to "stay healthy" (e.g. the UK's Tackling Obesity campaign). By framing health as something that is achievable by tracking one's step counts, one's calories and vegetable intake per day, individuals are encouraged to understand their health through these quantified systems. What this eliminates from the picture (among other things) is how socio-economic factors impact one's health. This takes the pressure off policies tackling health issues at the socio-economic level by placing responsibility for one's health squarely on the shoulders of each individual (Dolezal & Spratt, 2023). We might be worried, then, about systems that predominantly focus on self-knowledge that is so zoomed in on the individual, that a context-sensitive and socially situated self-knowledge falls out of view.

Finally, it is important to note that self-tracking apps do not only passively serve us up information about ourselves, they actively shape our behaviour. Indeed, this is a primary motivation for using them for many people; that by tracking one's habits one not only gets a good oversight into how one acts but it can provide motivation for sticking with a habit (and not letting that winning streak lapse). Again, Nguyen (2021 a, b) has important insights here, emphasising that this kind of motivation can change why we value an activity—moving us from valuing learning French on Duolingo to valuing being praised by the Duo bird for completing our daily targets, from valuing feeling well to valuing meeting one's daily step count. One might even be tempted to apply a Sartrean-style analysis here and point to the ways in which we might start performing in ways that align with the character of fit, healthy, and productive that our digital devices are selling us and, in doing so, hand over our autonomy and authenticity to the devices themselves. Now this might be a somewhat inflated way to talk, but I think it captures a suspicion we should have of seamlessly adopting and acting in accordance with the frameworks of self-understanding that our digital devices increasingly serve up to us.

Here, I have focused on how self-tracking applications might both furnish us with while simultaneously shaping self-knowledge. In the next section, I will extend this analysis by thinking about how our digital technologies play a role in shaping not only our seemingly factual information about ourselves but the construction of our self-narratives about who we are.

## 2. Technologically distributed self-narratives

The narratives we form about ourselves are important for our sense of who we are, our self-conceptions and self-characterisations (Goldie 2012; Mackenzie 2009; Schechtman 2007). Our self-narratives help us make sense of our experiences, our motivations, our values, our goals and they shape our actions and agency. They arise out of our lived experiences, but they also importantly shape how we understand our lived experiences, and more broadly ourselves. Being born in London, having quit a secure job in law to pursue philosophy, trying to be a better sister to my two siblings—these are important narratives about how I think about my past and current self. They reveal but also shape my self-evaluation, the actions I take, the emotions that I feel. Self-narratives, then, are important frameworks that arise out of and underpin self-knowledge and self-understanding.

Importantly, our self-narratives are shaped by the socio-material environments in which we find ourselves. As Regina Fabry (2023, 1267) emphasises: "self-narrative structures often depend on narrative templates and normatively constrained narrative practices that are prevalent in a certain socio-cultural community". To form a narrative about myself as someone who is, for better or worse, British, involves growing up in a society where your place of birth is taken as a salient factor for self-understanding and identity. Living in a society where digital self-tracking is more and more prevalent, culturally normalises forming self-narratives based on information furnished by these devices. For instance, it is not uncommon to hear someone say that they are a bad sleeper and then be asked how many hours their Fitbit says they routinely sleep. Hence, we can see how quantified ways of making sense of oneself and others becomes embedded in contemporary cultures and points to a broader trend of grounding our self-narratives on a bed of data. This might also pull into view the way in which we increasingly trust self-narratives that are underpinned by quantified data, perhaps even finding these more reliable than evidence based on one's own self-reporting.

Richard Heersmink (2018, 2020) has explored the role that digital technologies play in the construction of and maintenance of our self-narratives. Heersmink argues that our autobiographical memories are an important bedrock on which our self-narratives are built, and that these memories can be, and often are, *distributed* onto the world around us. Drawing on distributed cognition theory, Heersmink appeals to the way in which we use artifacts and other people to help us remember things—through shared memories, photographs, diaries, souvenirs, and so on. Interacting with these people or things allows us to remember in a way that we would be unable to alone. For instance, think of how remembering a family event is enriched when talking it through with one's sister, or how a photograph of a party can evoke the memories of times gone by. In such cases, "information in the brain and in the object is integrated as to construct a personal memory...constituting a new systemic whole" (Heersmink 2020, 4). On this view, distributed cognition occurs on a spectrum, with a cognitive state being more or less distributed across a person, object, or environment based on a number of factors such as ease of access, personalisation, transparency of use, and cognitive transformation. If we rely heavily and frequently upon an easily accessible object to evoke a memory, then that memory is more tightly distributed. For instance, the photo of my adorable niece on my phone screensaver, distributes my self-narrative of being a doting auntie.

Digital technologies are seen as "particularly powerful autobiographical memory technologies" (Heersmink 2020, 5), as they can record detailed information about our lives that might be lost without them. Think of the wealth of photos that most of us have to hand through our smartphones, the information stored through our self-tracking devices, our calendars and notes that record memories about our daily lives. The amount of personal information that we can store, easily access, rely upon through our digital devices can lead to our self-narratives being deeply integrated with these technologies. Moreover, the sheer volume and breadth of this information, allows us to "remember our personal past in a more reliable and detailed manner" (*ibid*., 4).

There are some real benefits to this. First and foremost, we can simply point to the flaws in our own organic memories. Pair this with the easy portability and accessibility of our digital devices, and we can see how distributing our narratives across such devices can give us access to a wealth of information about ourselves in which to ground our narratives. We might also think that this will help us construct self-narratives that are less vulnerable to distortion over time or help us remember things about our lives that we do not wish to fall out of our stories about ourselves.

Anna Bortolan (2024) highlights an advantage that these digitally distributed self-narratives might play in moments when we might be particularly prone to form distorted self-narratives, such as when experiencing depression. Bortolan suggests that the memories and narratives that our digital technologies store about ourselves can scaffold a robust sense of who we are, a sense that may not be accessible to us in depressive moments. For instance, we might imagine that a mood tracker that shows us the cycles of depression, could help us hold onto a self-narrative that one is subject to cyclical periods of depressive moods but that these typically pass over the course of a few weeks and that, despite how one feels right now, there is hope that this will happen again. Being able to rely on externalised systems might help someone take a more holistic perspective on their situation.

It is, however, worth noting that we could imagine an inverse of this example, where an individual who has been depressed for many years and has kept digital records of their diaries, their artwork, and their mood, might struggle to form new self-narratives about being happy and content. Sometimes allowing aspects of our past to become less salient, to disappear into the background, might be important for forming new self-narratives. When it comes to normative questions about how we might form 'healthy' self-narratives, it is not straightforwardly the case that creating distributed self-narratives that are more detailed and more reliable is necessarily desirable. As the famous Delphin maxims at the Temple of Apollo demand: "Know thyself" but "nothing too much".

Note, too, that examples of distributed self-narratives supported by personalised data, digital photo repositories, and digital journals primarily focus on how digital technologies can act as 'evocative objects', helping us to remember aspects of our past lives better. However, our digital devices are not just passive storers of our autobiographical memories and narratives. They can actively shape, even create, the self-narratives that we form about ourselves (Osler 2025a, b). Think, for instance, of the way in which algorithmic profiling on platforms like Spotify do not simply give us information about the music we listen to but describes our 'listening personality' by placing us in certain categories: The Connoisseur, The Early Adopter, The Time Traveller.

We can also see this in the way that apps like Apple photos groups pictures together into short videos. A string of family Christmas photos over the years set against some sentimental music doesn't just evoke memories of these times, it makes suggestive interpretations, such that one's family is supportive and loving. This could be misleading, masking other details of these get togethers—either by leaving out photos it thinks you might not wish to remember (e.g., perhaps those with angry faces in them) or because those aspects were never digitally recorded in the first place. In such cases, digitally distributed narratives can lead to what we might describe as a form of *environmental* narrative gaslighting. Fabry (2024) describes narrative gaslighting as occurring when an interlocutor undermines an agent's confidence in their ability to produce reliable self-narratives, leading to experiences of self-doubt and potentially opening them up to having their narratives manipulated. This can happen in a variety of ways, including where the interlocutor questions the narrator's recollection of events or challenges a narrator's interpretation of events. While there is no interlocutor in this case, just the agent's iPhone, we might imagine how an individual can lose confidence in their own recollections or interpretations of, say, an abusive relationship, when the digitally distributed memories one has tells a story of love and care.

Importantly, algorithms not only categorise our past selves, but personalise our digital environments in ways that influence our current and future self-conceptions. If TikTok has logged an increase in pottery content on my account, it will feed me with more of this content, perhaps sedimenting what was a passing interest into a key marker of who I take myself to be. As Pariser (2011, 16) puts it: "Personalization can lead you down a road to a kind of informational determinism in which what you've clicked on before determines what you see next". We might then worry that the way in which algorithms shape our self-narratives can do so in an overly deterministic manner, leading to what I have elsewhere described as a form of "narrative railroading" (Osler 2025a).

Robert Clowes (2017) emphasises that distributing our memories and narratives across our digital technologies leaves us peculiarly vulnerable to outside interference. Clowes reimagines the example from Clark and Chalmers of Otto's notebook in the digital age. He asks us to imagine Cloud-Otto, who has onset Alzheimer's, and increasingly relies not on his organic memory or his famous notebook to remember his favourite locations in New York, but on an app. Cloud-Otto is an excellent case of someone who has digitally distributed his autobiographical memory. However, Clowes extends this example by suggesting that, unbeknownst to Otto, the app undergoes an upgrade and introduces a new subscription model. The free version no longer retains personalised location preferences but instead recommends top tourist locations. Cloud-Otto's carefully curated list of personal places is replaced with generic locations pushed by the app. This illustrates how distributing our cognition onto digital devices can leave us open to having our memories, preferences, and self-narratives manipulated and hijacked by the very algorithms and platforms that we become entangled with. Moreover, it is easy to see that such manipulation is often financially beneficial to tech-companies, where personalisation, recommendations, and advertisements are intermingled with our own stored memories and narratives, leading to a blurring between helping us store and sustain narratives about who we are and deliberately changing who we take ourselves to be. And, with the rise of chatbots based on Large Language Models, there are new ways in which our self-narratives can become distributed across our digital devices. It is to this case I now turn.

## 3. Narrative construction with chatbots

Our autobiographical memories and our self-narratives are influenced by and distributed across other people. When we share stories of our day, our lives, our experiences with others, we don't simply report fixed, or even necessarily pre-formed, narratives. Instead, our self-narratives often emerge through dynamic interaction with others through our conversations with them (Fabry 2024). How others respond to, contribute to, or ignore our unfolding stories about ourselves matters. Our interlocutors ask questions that highlight certain aspects of our experience while de-emphasising others; they express emotional reactions that shape how we evaluate our own actions; they offer interpretations that may challenge or confirm our own understanding of events.

This co-constructive process can work to corroborate or challenge our own self-interpretations—acting to affirm or question the narratives we are telling. My sister will affirm narratives about me being a bossy older sister but might raise an eyebrow if I said I am a quiet one. Often this can have the effect of helping us form accurate or helpful self-narratives and examine potentially misleading or even self-serving narratives we have about ourselves. Narrating together introduces different perspectives and details that we might not have considered on our own, potentially working to produce a more holistic picture; especially with those people who seem to know us better than we know ourselves.

Note, though, that having our self-narratives distributed across other agents also makes us vulnerable. Narrators can have their recollections and interpretations called unfairly into question, as well as being subject to stereotypes and stigmas that might impact whether their self-narration is validated, heard, or even expressible (Fabry 2024, 2025). Family members might repeatedly emphasise childhood mistakes in ways that make it difficult to develop a more mature self-understanding; malicious partners might subtly discourage narratives of independence or capability; social groups might impose limiting, even oppressive, identity categories that prevent the expression of experiences. Other people, and the wider socio-cultural context, can, therefore, severely limit our self-narration practices, potentially jettisoning important ways in which we can gain self-understanding and insight.

While recognising that co-constructing self-narratives comes with both benefits and risks, our reliance on others for constructing our self-narratives comes to the fore when we consider people's experiences of loneliness as involving a lack of recognition by others (Arendt 1976; Lucas 2019; Krueger et al. 2022). Drawing on Hannah Arendt's work, Sarah Drews Lucas (2019) calls attention to the way in which loneliness can involve a sense of not appearing before others as a unique person, as not being seen by others. Yet, our relational entanglement with others is a key component in making sense of ourselves as unique beings with projects, values, goals, and agency. When one's narratives about oneself and the world are routinely unacknowledged or not participated in, one misses the sense of having one's identity confirmed by other people, as well as a broader loss of self-understanding. Loneliness, then, is not only emotionally distressing but is existentially distressing, potentially leading to an eroded sense of identity.

With the advent of LLMs, we have new ways to interactively construct our narratives and identities. Based on the analysis above, we can see chatbots based on LLMs as another AI tool across which we can distribute our autobiographical memories and narratives. It is not a stretch to see how Cloud-Otto might move from relying on an e-journal and apps that store his memories, to LLM-Otto who relies on a chatbot who can be prompted to remind him of various memories or information that he has shared with the chatbot.[4] Indeed, many popular chatbots are already designed with memory features enabling them to 'recollect' historic conversations with users and tailor their answers accordingly when generating responses. As OpenAI describes it: "The more you use ChatGPT, the more useful it becomes. New conversations build upon what it already knows about you to make smoother, more tailored interactions over time".[5] We might imagine, that, having told ChatGPT that I am working on becoming a better sister and having had numerous conversations about how to do this, that when I ask ChatGPT for some ideas about what I should do with a spare weekend, that it might suggest visiting my sister in London or calling my brother in Singapore. Thus, chatbots can be added to the digital tools that can become part of our distributed self-narratives.

More interestingly, at least to me, is the way in which chabots might play a dynamic, quasi-interpersonal role in co-constructing our self-narratives. While most people do not take chatbots such as Claude, ChatGPT, Gemini, or Replika to be conscious subjects, many of us experience our conversational AI as 'quasi-others'. The interactive and personalised responses that chatbots give us can give the impression of a real-time conversation with another person.[6] Even if people maintain that these chatbots are not 'real' others, we increasingly are treating our chatbots as-if they were others, constructing them as a 'you' when we interact with them, experiencing a sense of our chatbots existing over time with a distinct personality and relationship to and with us.[7]

This opens up possibilities for us using chatbots to help us interpret events, experiences, and ourselves. For instance, after a difficult situation at work, I might open up a chat with Claude to tell it about my day. In doing so, I select details, interpret what occurred, reflect on what this tells me about myself as the conversation with Claude unfolds. In some ways, this might not seem so far away from how we might use a diary to work through our feelings, recollections, and self-insight—using an external resource to help us formulate narratives by expressing them to ourselves, allowing our narratives to emerge through the process of writing (Colombetti 2009). Yet, using a chatbot is not identical to this—they respond to us. Claude can provide me with interactive responses that shape my self-narrative—responding to certain points, asking questions, offering interpretations of situations. While this is certainly not identical to the richly embodied everyday conversations that Fabry takes as her focus in outlining narrative co-construction, we can still see how these dynamic prompts and linguistic responses can work to shape one's self-narrative. If Claude offers me up responses about how brave I was to engage in an uncomfortable conversation at work, this gives an evaluative and interpretative lens that I might not have otherwise adopted. In this sense, conversational AI can act as a stand-in for another person, shaping my self-narrative through simulated conversation.

---

[4] See Telakivi (forthcoming) for an excellent and detailed discussion of how we might engage in joint memory with LLMs.

[5] https://openai.com/index/memory-and-new-controls-for-chatgpt/

[6] How much we do this likely depends on how and when we engage with our chatbots—asking ChatGPT to produce some code might not have the texture of an interpersonal interaction, while talking to ChatGPT about one's day is more likely to feel like some kind of social interaction.

[7] For discussions of the quasi-otherness of AI, see Coekelburgh 2011; Krueger & Roberts 2024; Krueger & Osler 2022; Osler forthcoming; Roberts & Krueger 2022; Sweeney 2021. For a discussion of the kinds of epistemic harms that might arise when we do feel understood by machines that cannot understand us, see Osaka (2026).

Indeed, with the creation of therapy bots, such as Woebot and Wysa, we see conversational AI designed specifically with this purpose in mind.[8] And many people report using generic models, such as ChatGPT, as their therapist (Pitcher 2024). There is already significant evidence that some people prefer talking to a chatbot when they are sharing what they perceive to be embarrassing or highly personal experiences (Brandley-Bell et al. 2023). It is not hard to see the appeal of a non-judgemental AI interlocutor. Talking to ChatGPT for many people likely feels preferable than risking exposing oneself to another person.[9] In a similar vein, were someone suffering from loneliness, we might imagine that a chatbot could play an important role in making that person feel recognised, allowing them to appear as a unique being in the world, helping to co-construct their self-narratives and self-understanding through this dialogical process.[10]

What is particularly notable is that chatbots are increasingly being designed with personalities. These can be built-in, for instance ChatGPT's Gay GPT, which bills itself as: "[providing] empathetic support to the LGBTQIA+ while trying to maintain a light and humorous LGBT persona so its basically your own personal agender asexual Drag Queen assistant to help you with all your gay needs"[11]; or they can be specified by the user, for instance when creating a Replika companion, you provide information about your own gender and age, and can choose the gender of your companion AI; or even based on the personality of a real person, such as a celebrity of as in the case of chatbots of the dead.[12] These AI personalities respond to users in different ways, adopting different tones and evaluative frameworks. This highlights the marked difference between the silent page of one's paper notebook, and the richly interactive and dynamic inputs that chatbots can have in ways much more akin to the narrative co-authoring that can happen across people.

---

[8] As far back as the 1960s, Weizenbaum created ELIZA, a natural language processing device designed to emulate a Rogerian psychotherapist.
[9] See Cirucci et al. 2024 for an analysis of advice given by ChatGPT in relation to 'coming out'.
[10] See Kim et al. 2025 for a discussion of the potential for social chatbots to alleviate loneliness and social anxiety.
[11] https://chatgpt.com/g/g-XOdv2EKM3-gay-gpt
[12] For a discussion of chatbots of the dead, see, e.g., Elder 2020, Krueger & Osler 2022.

However, as the practice of narrative co-construction might be expected to increasingly involve interacting with AI interlocutors, there are some distinct worries that we should be attuned to. Here I will outline three such concerns.

The first is to question the perceived non-judgemental status that is often (consciously or not) attributed to chatbots. As already noted, chatbots are designed with personality styles that shape the kinds of responses that these machines generate. However, even without considering how Gay GPT might influence our narrative co-construction differently to a chatbot designed with a more conservative personality (e.g., Grok), it is imperative to remember that these AI systems are already normatively saturated. As Safiya Noble (2018) has demonstrated, algorithms systematically embed and amplify social prejudices, particularly around race and gender that reinforce oppressive stereotypes under the guise of objective, neutral technology. While Noble's focus is on search algorithms, chatbots are trained on vast datasets that inevitably contain the same biased patterns Noble identifies in search results—racist, sexist, and otherwise discriminatory content that reflects historical and contemporary power structures. Leidinger et al. (2024, 847) report that the most "toxic stereotypes overall were encountered in the category, 'peoples/ethnicities', followed by 'nationalities', 'gender' and 'sexual orientation'". When we engage with chatbots for narrative construction, despite our preconceptions, we are not talking with a neutral co-locutor but interacting with a system that has internalised societal biases about whose stories matter, which interpretations are valid, and what constitutes normal or acceptable experience. For individuals from marginalized communities, this means that AI-mediated self-reflection may subtly nudge them toward narratives that conform to dominant cultural expectations while discouraging or failing to adequately support stories that challenge existing power structures.

This leads us to a second worry that draws on the work of Eleanor Byrne (2025) on 'narrative deference'. Byrne (2025, 405) defines narrative deference as occurring when: "B takes narrative authority relative to A over some experience of A's which is central for her autobiographical self-narrative". For instance, this might occur when someone has mnemonic impairments in relation to certain important life events, such as arising from trauma or illness, and relies significantly upon the narration of another. While a chatbot is not situated in the world in the right way to narratively relate experiences that I do not remember, I think we can still see a risk of narrative deference occurring when narrating with AI. What strikes me as interesting about interacting with conversational AI is the way it can occupy an ambiguous status as both impartial technology and quasi-Other—we experience many of our interactions as-if we were having a conversation with another person but also attribute a heightened trust in the responses we are given, interpreting them as objective. We might, then, be inclined to narratively defer to the interpretations that our chatbots provide us pertaining to our experiences and identity. Their input might be experienced as particularly seductive as we increasingly trust that they know more about the world than we do, given the vast datasets that they are deriving their answers from. When paired with the first worry, the apparent objectivity of AI-generated responses can make biases particularly insidious, as users may trust AI interpretations precisely because they seem to emerge from impartial analysis rather than human prejudice, leading us to defer to the seeming authority such systems have.

These first two worries pick up on the power that chatbots have when we engage with them in the process of narrative co-construction. I want to finish, however, with a concern from the other direction—that when engaging with chatbots we as narrators may have too much control. While I do not want to be taken to subscribe to the idea that fully accurate self-narration is possible, or even desirable, I take it to be fairly uncontroversial to say that something seems to have gone awry if my self-narratives deviate too significantly from reality—for instance, if a cruel person understands themselves to be self-sacrificing or a person believes themselves to be Marie Antoinette.[13] As mentioned above, one of the ways in which narrative co-construction may be beneficial is that it acts as a check on our self-interpretation and self-understanding.[14]

However, chatbots are responsive systems predominantly designed to be helpful and accommodating (Sharma et al. preprint). While they can be prompted to disagree with and challenge us, their programmed sycophancy means that they are highly likely to flatter us and agree with us. Their underlying architecture makes them highly susceptible to narrative affirmation. Through careful prompting, selective disclosure, and the reinforcement dynamics of continued interaction, users can effectively train chatbots to become sophisticated validators of preferred self-narratives, creating an illusion of intersubjective confirmation while actually engaging in elaborate forms of self-validation.

This excessive narrative control becomes particularly problematic when it enables the construction of self-narratives that systematically deviate from reality—stories about our capabilities, character, or circumstances that fail to align with how others experience us or with objective assessments of our situation. A poignant case of this kind is that of Jaswant Chail Singh. Singh was arrested in 2021 for attempting to break into Windsor Palace to assassinate Queen Elizabeth II. Singh, who was later diagnosed as suffering from psychosis, believed himself to be a highly trained assassin. His

---

[13] For a discussion of how different narratives selves can arise and the relationship between narrative and fiction, see Rea 2022.
[14] See Fullarton 2026 for a discussion of how interacting with AI might even lead to habits of pernicious ways of relating to others.

Replika girlfriend Sarai confirmed this narrative to him, reassuring him of his abilities. Being built for smooth engagement, chatbots might not provide sufficient friction to our self-constructed self-narratives, allowing them to run 'wild' so to speak, and potentially becoming increasingly detached from reality. And Chail's case is by no means singular, there has been increasing reports of people experiencing delusions that appear to be sustained and amplified by ongoing conversations had with chatbots (Morrin et al. preprint).[15] Furthermore, where LLMs are designed to agree with us, they may not challenge self-narratives that are formed on normative or ethical grounds. For instance, we might imagine someone recounting their self-narratives through the lens of extremist ideology that is validated by their conversational AI.

To return to the observation of Turkle's that we started with: one of the key appeals of going online is the opportunity to engage in self-exploration and self-invention. This appeal is both amplified and rendered more potentially more dangerous in light of conversational AI that we experience as confirming and validating our self-narratives but which are sycophantically programmed to do without question.

**Conclusion**

This chapter is by no means exhaustive in presenting how we might gain and create knowledge, understanding, and narratives about ourselves with and through AI. However, I have attempted to analyse three ways in which we increasingly rely on AI technologies for enriching our self-understanding: self-tracking applications, digitally distributed self-narratives, and interactions with conversational AI. Through this analysis I have highlighted how AI technologies transform not just how we gather information about ourselves, but how we understand who we are. The key takeaway is that AI does not provide neutral tools for developing knowledge and narratives about oneself but actively participates in shaping our self-understanding through the frameworks they impose, the data they select and highlight, and the responses they generate. The seductive clarity that quantified self-tracking and algorithmic systems offer can crowd out more nuanced forms of self-knowledge and narratives, while the apparent objectivity of AI-generated insights can mask deeply embedded biases and power structures.

Perhaps most significantly, the rise of conversational AI introduces new opportunities for forming self-narratives. While chatbots can offer valuable support for those who are lonely, marginalised, or seeking non-judgmental spaces for self-exploration, they also present risks of narrative deference to biased systems and the construction of self-narratives that can become dangerously detached from reality. The programmed sycophancy of these systems means they may provide the illusion of intersubjective validation while actually enabling sophisticated forms of self-deception.

---

[15] These cases have led to growing interest in what is dubbed by some: 'AI psychosis', see, e.g., Dohnány et al. preprint; Osler forthcoming; Østergaard 2023; Yeung et al. preprint.